\documentclass[letterpaper, 10 pt, conference]{ieeeconf}
\usepackage{amssymb}
\usepackage{amsmath}
\usepackage{graphicx}
\usepackage{caption}
\usepackage{subcaption}
\usepackage{cleveref}

\IEEEoverridecommandlockouts                              
\newtheorem{theorem}{Theorem}

\newtheorem{lemma}[theorem]{Lemma}

\newtheorem{proposition}[theorem]{Proposition}
\newtheorem{remark}[theorem]{Remark}
\newtheorem{assumption}[theorem]{Assumption}
\begin{document}

\title{{\LARGE \textbf{Robust Moment Closure Method for the Chemical Master
Equation}}}
\author{Mohammad Naghnaeian and Domitilla Del Vecchio \thanks{%
M. Naghnaeian is a postdoctoral associate with the Mechanical Engineering
Department, Massachusetts Institute of Technology, Cambridge, MA, USA \texttt%
{\small mongh@mit.edu}} \thanks{%
D. Del Vecchio is with the Mechanical Engineering Department, Massachusetts
Institute of Technology, Cambridge, MA, USA \texttt{\small ddv@mit.edu}} 
\thanks{%
This work was supported by the Air Force Office of Scientific Research under
grant FA9550-14-1-0060}}
\maketitle

\begin{abstract}
The Chemical Master Equation (CME) is used to stochastically model
biochemical reaction networks, under the Markovian assumption. The low-order
statistical moments induced by the CME are often the key quantities that one
is interested in. However, in most cases, the moments equation is not
closed; in the sense that the first $n$ moments depend on the higher order
moments, for any positive integer $n$. In this paper, we develop a moment
closure technique in which the higher order moments are approximated by an
affine function of the lower order moments. We refer to such functions as
the affine Moment Closure Functions (MCF) and prove that they are optimal in
the worst-case context, in which no a priori information on the probability
distribution is available. Furthermore, we cast the problem of finding the
optimal affine MCF as a linear program, which is tractable. We utilize the
affine MCFs to derive a finite dimensional linear system that approximates
the low-order moments. We quantify the approximation error in terms of the $%
l_{\infty }$ induced norm of some linear system. Our results can be
effectively used to approximate the low-order moments and characterize the
noise properties of the biochemical network under study.
\end{abstract}

\thispagestyle{empty} \pagestyle{empty}

\section{Introduction}

Biomolecular reaction networks are mostly studied in two frameworks,
deterministic or stochastic \cite{oppenheim1969stochastic}. In the former,
the system is modeled by a set of Ordinary Differential Equations (ODEs)
whose states represent the concentration of the species. Such models have
proved to be useful in explaining and predicting the behavior of the system
especially in the high concentration regime. They, however, fail to
accurately explain the characteristics of the system in the low
concentration regime \cite{kurtz1972relationship}. In fact, when the number
of molecules in the network is low, the inherent randomness in the
interactions and the discreteness of the system's state play an important
role towards the overall behavior \cite{rao2002control}. This necessitates
the use of stochastic models.

In stochastic framework, the Chemical Master Equation (CME) is used to model
biochemical reaction networks \cite{del2015biomolecular}. It is a popular
modeling framework in the systems biology community, in which it has been
widely employed to study the impact of intrinsic noise on a network's
behavior and to capture the behavior of networks characterized by low
molecule counts \cite{matheson1975stochastic}. Although the CME is a linear
system, its explicit solution cannot be obtained, in general. This is due to
the fact that, except in very idealistic situations, the dimension of the
CME is very large or often infinite. A reasonable approach to cope with this
curse of dimensionality is to study the statistical moments.

Low-order statistical moments, particularly the first and second moments,
are often the key quantities that one is interested in as they provide
indication on standard noise quantifications, such as the coefficient of
variation. One difficulty that arises in this approach is that, in most
cases, the moments equation induced by the CME is not closed \cite%
{sotiropoulos2011analytical}; in the sense that the first $n$ moments depend
on the higher order moments, for any positive integer $n$. This is
challenging since the low-order statistical moments of the system cannot be
studied without knowing the higher-order moments due to this coupling. For
analysis and simulation purposes, one can close the system of moments by
approximating the higher-order moments. In the literature, such a procedure
is referred to as moment closure. Any moment closure technique consist of
two steps \cite{sotiropoulos2011analytical}:

\begin{description}
\item[(a)] The statistical moments higher than $n$ are approximated as a
function (possibly nonlinear) of the first $n$ moments. This function is
called the Moment Closure Function (MCF).

\item[(b)] The high-order moments in the low-order moments equations are
replaced by the MCF. This results in a closed system for the low-order
moments.
\end{description}

There are various moment closure methods proposed in the literature. Most of
them assume an underlying probability distribution. For example, in \cite%
{whittle1957use}, \cite{singh2006lognormal}, and \cite{krishnarajah2005novel}
the probability distribution is assumed to be normal, log-normal, and
beta-binomial, respectively. There are also techniques that are not
distribution based. For instance, \cite{matis2002interacting} uses cumulant
truncation and \cite{singh2011approximate} uses the derivative-matching.
Upon utilizing any moment closure method, the resulting closed system serves
as an approximation to the low-order moments. Hence, it is important to
quantify the error of this approximation. To the best of authors knowledge,
no such quantifications are available in the literature. Accordingly, the
development of a moment closure method with quantifiable error bounds deems
necessary and this is what this paper aims to address.

In this paper, we develop the Robust Moment Closure (RMC) method for which
we can exactly quantify the approximation error. In this method the higher
order moments are approximated by an affine function of the lower order
moments. We mathematically prove that affine MCFs are optimal in the
worst-case context, in which we do not have a priori information on the
probability distribution. In this case, no (possibly nonlinear) MCF can
outperform the affine ones. We show that finding the optimal affine MCF is a
Linear Program (LP) and hence tractable \cite{luenberger1973introduction}.
Consequently, utilizing the affine MCFs, we derive a set ODEs of finite
dimension that approximates the time evolution of the lower order moments.
Furthermore, we quantify the error in this approximation in terms of the $%
l_{\infty }$ induced norm of some linear system. Our results allow for the
explicit simulation and analytical computation of approximate moments, which
can be used to characterize the noise properties of the biomolecular
reaction networks. The proofs of the
results are given in the Appendix.

\section{Preliminaries}

The following notations are used throughout this paper: $\mathbb{Z}_{\geq 0}$
and $\mathbb{R}_{\geq 0}$ is the set of nonnegative integers and real
numbers, respectively. For a positive integer $n$, $\mathbb{Z}_{\geq 0}^{n}$
($\mathbb{R}_{\geq 0}^{n}$) denotes the set of $n$-dimensional vectors with
entries in $\mathbb{Z}_{\geq 0}$ ($\mathbb{R}_{\geq 0}^{n}$). Given an $n$%
-dimensional vector $X=\left[ x_{1},x_{2},...,x_{n}\right] ^{T}$ and a
nonnegative integer $I$, define $\Psi _{I}\left( X\right) $ to be the vector
composed of entries of the form $x_{1}^{k_{1}}x_{2}^{k_{2}}...x_{n}^{k_{n}}$
where $k_{i}\in \mathbb{Z}_{\geq 0}$, for $i=1,2,...,n$, and $%
\sum_{i=1}^{n}k_{i}=I$. For example, for $X=\left[ x_{1},x_{2}\right] ^{T}$,
we have%
\begin{eqnarray*}
\Psi _{1}\left( X\right) &=&\left[ x_{1},x_{2}\right] ^{T},\Psi _{2}\left(
X\right) =\left[ x_{1}^{2},x_{1}x_{2},x_{2}^{2}\right] ^{T}, \\
\Psi _{3}\left( X\right) &=&\left[
x_{1}^{3},x_{1}^{2}x_{2},x_{1}x_{2}^{2},x_{2}^{3}\right] ^{T}.
\end{eqnarray*}%
Also, define $\Psi _{0}\left( X\right) =1$. Given a positive integer $n$,
define the vector 
\begin{equation}
\bar{\Psi}_{n}\left( X\right) =\left[ \Psi _{1}^{T}\left( X\right) ,\Psi
_{2}^{T}\left( X\right) ,...,\Psi _{n}^{T}\left( X\right) \right] ^{T},
\label{eq:ci_bar}
\end{equation}%
and, for $i=1,2,...,n$, let $c_{i}$ be a matrix whose multiplication with $%
\bar{\Psi}_{n}\left( X\right) $ isolates $\Psi _{i}\left( X\right) $, i.e. 
\begin{equation}
\Psi _{i}\left( X\right) =c_{i}\left[ \Psi _{1}^{T}\left( X\right) ,\Psi
_{2}^{T}\left( X\right) ,...,\Psi _{n}^{T}\left( X\right) \right] ^{T}.
\label{eq:ci}
\end{equation}%
The $l_{\infty }$ and $l_{1}$ norms of a vector $X=\left[
x_{1},x_{2},...,x_{n}\right] ^{T}$ are defined as $\left\Vert X\right\Vert
_{l_{\infty }}:=\max_{i}\left\vert x_{i}\right\vert $ and $\left\Vert
X\right\Vert _{l_{1}}=\sum_{i=1}^{n}\left\vert x_{i}\right\vert $. We use $%
\left\Vert X\right\Vert $ without any subscript to mean $\left\Vert
X\right\Vert _{l_{\infty }}$. A vector $P\in \mathbb{R}_{\geq 0}^{p}$ is
called a probability vector if $\left\Vert P\right\Vert _{l_{1}}=1$. The set
of all probability vectors with dimension $p$ is denoted by $\mathbb{P}^{p}$%
. We omit the superscript $p$ when the dimension is irrelevant or obvious
from the context. Given a matrix $M=\left[ m_{ij}\right] \in \mathbb{R}%
^{m\times n}$, by $\mathcal{R}\left[ M\right] _{i}$ and $\mathcal{C}\left[ M%
\right] _{j}$ we mean the $i^{th}$ row and $j^{th}$column of $M$,
respectively. That is,%
\begin{eqnarray*}
\mathcal{R}\left[ M\right] _{i} &=&\left[ 
\begin{array}{cccc}
m_{i1} & m_{i2} & \cdots & m_{in}%
\end{array}%
\right] , \\
\mathcal{C}\left[ M\right] _{j} &=&\left[ 
\begin{array}{cccc}
m_{1j} & m_{2j} & \cdots & m_{mj}%
\end{array}%
\right] ^{T},
\end{eqnarray*}%
for $i=1,2,...,m$ and $j=1,2,...,n$. The $l_{1}$, $l_{\infty }$, and $l_{1}$
to $l_{\infty }$ induced norms of $M$ are defined as $\left\Vert
M\right\Vert _{l_{1}-ind}=\max_{j}\sum_{i=1}^{m}\left\vert m_{ij}\right\vert 
$, $\left\Vert M\right\Vert _{l_{\infty
}-ind}=\max_{i}\sum_{j=1}^{n}\left\vert m_{ij}\right\vert $, and $\left\Vert
M\right\Vert _{l_{1}-l_{\infty }}=\max_{i,j}\left\vert m_{ij}\right\vert $.
Furthermore, the null space of $M$ and its perpendicular complement (perp)
are denoted respectively by $\mathcal{N}\left( M\right) $ and $\mathcal{N}%
^{\bot }\left( M\right) $ and defined as%
\begin{eqnarray*}
\mathcal{N}\left( M\right) &=&\left\{ x:Mx=0\right\} , \\
\mathcal{N}^{\bot }\left( M\right) &=&\left\{ y:y^{T}x=0,\forall x\in 
\mathcal{N}\left( M\right) \right\} .
\end{eqnarray*}%
Also, define $N\left( M\right) $ and $N^{\bot }\left( M\right) $ to be
matrices whose columns form orthonormal basis for $\mathcal{N}\left(
M\right) $ and $\mathcal{N}^{\bot }\left( M\right) $, respectively. The
following lemma holds:

\begin{lemma}
\label{lemma:01}Given a matrix $M=\left[ m_{ij}\right] \in \mathbb{R}%
^{m\times n}$, we have%
\begin{equation*}
\sup_{P\in \mathbb{P}}\left\Vert MP\right\Vert =\left\Vert M\right\Vert
_{l_{1}-l_{\infty }}.
\end{equation*}
\end{lemma}

Markov processes can be used to describe the dynamics of chemical reaction
networks. Each state of this Markov process represents the aggregated
molecule counts of the species. A transition from one state to another state
occurs when a chemical reaction fires and, as a result, the molecule counts
of species change. More precisely, suppose a reaction network with $q$
number of species and $J$ number of reactions. Let $s_{i}$, for $i=1,2,...,q$%
, be the count of each species and let $S=\left[ s_{1},s_{2},...,s_{q}\right]
^{T} $. Associated with each reaction $j\in \left\{ 1,2,..,J\right\} $,
there are a propensity function $a_{j}\left( t,S\right) $ and a
stoichiometry vector $\gamma _{j}$ defined as%
\begin{equation*}
\Pr \left( S\left( t+dt\right) =S\left( t\right) +\gamma _{j}|S\left(
t\right) \right) =a_{j}\left( t,S\left( t\right) \right) dt+O\left(
dt^{2}\right) ,
\end{equation*}%
with $\gamma _{j}$ is the change in species count upon firing of reaction $j$%
. In this case, for any $k\in \mathbb{Z}_{\geq 0}^{q}$, the probability
vector satisfies 
\begin{eqnarray}
&& \frac{d}{dt}\Pr \left( S\left( t\right) =k\right)=\sum_{j=1}^{J} \biggl\{ %
-a_{j}\left( t,k\right) \Pr \left( S\left( t\right)=k\right)  \notag \\
&& +a_{j}\left( t,k-\gamma _{j}\right) \Pr \left( S\left( t\right) =k-\gamma
_{j}\right)\biggr\}.  \label{eq:CME1}
\end{eqnarray}
This equation is referred to as the Chemical Master Equation \cite%
{van1983stochastic}\cite{gillespie1992rigorous}. Throughout this paper, we
make the following assumptions.

\begin{assumption}
\label{assumption:1}There exist nonnegative integers $U_{i}$ such that 
\begin{equation*}
0\leq s_{i}\leq U_{i},
\end{equation*}%
for $i=1,2,...,q$.
\end{assumption}

\begin{assumption}
\label{assumption:2}The propensity functions are polynomial in $S$ \cite%
{gillespie1976general}\cite{gillespie1992rigorous}. That is, for $%
j=1,2,...,J $,%
\begin{equation*}
a_{j}\left( t,S\right) =\sum_{i=0}^{l}\theta _{i}^{j}\left( t\right) \Psi
_{i}\left( S\right) ,
\end{equation*}%
for some $l\in \mathbb{Z}_{\geq 0}$, where $\theta _{i}^{j}\left( t\right) $%
's are matrices with appropriate dimensions.
\end{assumption}

Assumption \ref{assumption:1} states that we have an upper bound on the
number of molecules for each species. This assumption is readily satisfied
for species that are conserved in the biochemical reaction network, such as
DNA copy number or total protein concentrations in enzymatic reactions \cite%
{del2015biomolecular}. In the presence of species that are not conserved,
one can use the methods given e.g. in \cite{gupta2014scalable} to truncate
the system and find an upper bound on the species count such that the
truncated (finite dimensional) system is arbitrarily close to the infinite
dimensional CME. Regarding Assumption \ref{assumption:2}, we refer the
reader to \cite{gillespie1976general}, \cite{gillespie1992rigorous}, and 
\cite{gillespie2014validity} where the polynomial propensity functions are
derived under suitable conditions such as well-mixedness.

\section{Basic Setup}

Consider the CME given in (\ref{eq:CME1}) with Assumptions \ref{assumption:1}
and \ref{assumption:2}. This is a linear system of ODEs describing the time
evolution of the probability distribution vector of the underlying Markov
process. Based on Assumption \ref{assumption:1}, the CME is of order $p$,
where 
\begin{equation}
p:=\prod\limits_{i=1}^{q} \left( 1+U_{i}\right).  \label{eq:order}
\end{equation}%
Clearly, the order of a CME grows exponentially with respect to the number
of species present in the system. In most cases, the CME is a high
dimensional system and hence solving it is a computationally challenging
task. Thus, instead of directly solving the CME, one can consider the
low-order statistical moments. While the statistical moments of a
probability distribution are informative quantities to consider, they
contain less information than the distribution itself. Hence, intuitively,
one hopes for a less complex problem if only the moments are considered. In
the next proposition we derive the moments equation induced by the CME in (%
\ref{eq:CME1}).

We denote the $i^{th}$ moment of the random variable $S$ by $\mu _{i}$.
Recall that $\mu _{i}:=\mathbb{E}\left[ \Psi _{i}\left( S\right) \right]
=\sum_{k\in \mathbb{Z}^{q}}\Psi _{i}\left( k\right) \Pr \left( S=k\right) $,
where $\Psi _{i}\left( S\right) $ is a vector composed of the entries of the
form $s_{1}^{k_{1}}s_{2}^{k_{2}}...s_{q}^{k_{q}}$ with $%
k_{1}+k_{2}+...+k_{q}=i$. The following proposition holds (see e.g. \cite%
{sotiropoulos2011analytical} for the proof):

\begin{proposition}
\label{Prop:01}For the chemical master equation in (\ref{eq:CME1}) with
Assumptions \ref{assumption:1} and \ref{assumption:2}, for $i=1,2,...$, 
\begin{equation}
\frac{d}{dt}\mathbb{\mu }_{i}\left( t\right) =\beta _{i,0}\left( t\right)
+\sum_{n=1}^{i+l-1}\beta _{i,n}\mu _{n}\left( t\right),  \label{eq:03}
\end{equation}
with initial condition $\mu _{i}\left(0\right) =\Psi _{i}\left( S\left(
0\right) \right)$, for some properly defined matrices $\beta _{i,n}\left(
.\right)$ with appropriate dimension.
\end{proposition}

When $l>1$, the system of moments in (\ref{eq:03}) is not closed in the
sense that the lower-order moments depend on the higher-order moments. This
introduces a certain degree of complexity into the system. More precisely,
one cannot consider the low-order moments decoupled from the high-order
ones. Therefore, one needs to study the full system of moments including all
the moments up to order $p$, where $p$ is defined in \ref{eq:order} and
generally is very large as it scales exponentially with respect to the
number of species. Therefore, the full system of moments up to order $p$,
although closed, but is a high dimensional set of ODEs whose study is as
difficult as that of CME. Therefore, in the literature, there has been a
great deal of effort to approximate the higher-order moments, $\left( \mu
_{n+1},\mu _{n+2,}...,\mu _{n+l}\right) $, by a possibly nonlinear function
of low-order moments. This procedure is referred to as moment closure.
Unfortunately, the lack of error quantification prevails amongst the moment
closure methods. In the next section, we introduce the \textit{Robust Moment
Closure} technique for which we exactly quantify the error between the true
system (\ref{eq:03}) and the resulting closed system of moments.

\section{Robust Moment Closure}

Any moment closure method revolves around the idea of approximating the
higher order moments by a possibly nonlinear function of lower order
moments. This allows for closing the system of moments which in turn can be
more easily analyzed. In this section, we first discuss on the optimal MCF
in the worst-case setting; that is, when no a priori information on the
probability distribution is available. For the rest of this paper, we assume
that $l=2$ in (\ref{eq:03}). This assumption is made for two reasons. First,
any biochemical reaction, with more than two reactants , can be written as a
series of mono- or bi-molecular reactions that result in propensity
functions of order at most two \cite{gillespie2014validity}. Second, our
results can be easily extended to the case $l>3$ as remarked later.
Therefore, without loss of generality, we assume $l=2$ and obtain%
\begin{equation}
\frac{d}{dt}E_{n}\left( t\right) =A\left( t\right) E_{n}+b\left( t\right)
\mu _{n+1}+r\left( t\right) ;\text{ with }E_{n}\left( 0\right) \text{ given}
\label{eq:03'}
\end{equation}%
where $E_{n}:=\left[ \mu _{1}^{T},\mu _{2}^{T},...,\mu _{n}^{T}\right] ^{T}$
is the aggregation of all moments up to order $n$, and $A\left( t\right) $
and $b\left( t\right) $ are matrices with appropriate dimension \cite%
{sotiropoulos2011analytical}. Define matrices $H_{n}$ and $V_{n}$ such that 
\begin{eqnarray}
\mu _{n+1}\left( t\right) &=&H_{n}P\left( t\right) ,  \label{eq:H} \\
E_{n}\left( t\right) &=&V_{n}P\left( t\right) ,  \label{eq:V}
\end{eqnarray}%
where $P\in \mathbb{R}_{\geq 0}^{p}$ is the vector composed of entries $\Pr
\left( S=k\right) $ with $k\in \mathbb{Z}_{\geq 0}^{q}$ and $k\leq U$. For
example, for a one dimensional random variable $S$, 
\begin{eqnarray*}
H_{n} &=&\left[ 
\begin{array}{ccccc}
0 & 1^{n+1} & 2^{n+1} & \cdots & U^{n+1}%
\end{array}%
\right] , \\
V_{n} &=&\left[ 
\begin{array}{ccccc}
0 & 1 & 2 & \cdots & U \\ 
0 & 1^{2} & 2^{2} & \cdots & U^{2} \\ 
\vdots & \vdots &  &  &  \\ 
0 & 1^{n} & 2^{n} & \cdots & U^{n}%
\end{array}%
\right] .
\end{eqnarray*}

\begin{remark}
Notice that if $l>2$ in (\ref{eq:03}) the moments equation takes the form%
\begin{equation*}
\frac{d}{dt}E_{n}\left( t\right) =A\left( t\right) E_{n}+b\left( t\right) %
\left[ \mu _{n+1}^{T},...,\mu _{n+l-1}^{T}\right] ^{T}+r\left( t\right) .
\end{equation*}%
In this case, we modify the definition of $H_{n}$ given in (\ref{eq:H}). We
define $H_{n}^{l}$ as matrix such that%
\begin{equation*}
\left[ \mu _{n+1}^{T},\mu _{n+2}^{T},...,\mu _{n+l-1}^{T}\right]
^{T}=H_{n}^{l}P.
\end{equation*}%
Then, the results of this paper hold valid with $H_{n}$ replace by $%
H_{n}^{l} $. Hence, without loss of generality we assume $l=2$.
\end{remark}

Suppose that we are interested in closing the system of moments for the
first $n$ moments. To this end, we approximate $\mu _{n+1}$ by $\phi \left(
E_{n}\right) $, where $\phi \left( .\right) $ is some (possibly nonlinear)
function of the first $n$ moments. In this case, the closed system of
moments is given by 
\begin{equation}
\frac{d}{dt}\nu =A\nu +b\phi \left( v\right) +r;\text{ with }\nu \left(
0\right) =E_{n}\left( 0\right) ,  \label{eq:11}
\end{equation}%
which is analogous to (\ref{eq:03'}) with $\mu _{n+1}$ replaced by $\phi
\left( E_{n}\right) $. The function $\phi $ is the MCF and should be chosen
such that the error between $\mu _{n+1}$ and $\phi \left( E_{n}\right) $ is
minimized. This error is clearly a function of the probability vector and $%
\phi $. More precisely, define%
\begin{equation*}
\rho _{NL}\left( P,\phi \right) =\left\Vert \mu _{n+1}-\phi \left(
E_{n}\right) \right\Vert ,
\end{equation*}%
where the norm $\left\Vert .\right\Vert $ is taken to be the $l_{\infty }$
norm. Above, we have made the dependency of $\rho _{NL}$ on $\phi \left(
.\right) $ and the probability vector $P$ explicit; and the subscript $NL$
in $\rho _{NL}$ refers to the fact that $\phi $ can be a nonlinear function,
in general. Further, since the probability vector is not known, in the
Robust Moment Closure (RMC) technique, $\phi \left( .\right) $ is chosen
such that the worst-case error is minimized. This amounts to the following
min-max problem:%
\begin{eqnarray}
\rho _{NL}^{o} &=&\inf_{\phi }\sup_{P\in \mathbb{P}}\rho \left( P,\phi
\right)  \label{eq:05} \\
&=&\inf_{\phi }\sup_{P\in \mathbb{P}}\left\Vert H_{n}P-\phi \circ
V_{n}P\right\Vert ,
\end{eqnarray}%
where $P$ is restricted to the set of probability vectors, $P\geq 0$ and $%
\left\Vert P\right\Vert _{l_{1}}=1$. To solve this optimization problem,
notice that any $P\in \mathbb{P}$ can be uniquely written as%
\begin{equation}
P=N\left( V_{n}\right) x+N^{\bot }\left( V_{n}\right) y,  \label{eq:1}
\end{equation}%
for some $x\in \boldsymbol{R}^{p-r}$ and $y\in \boldsymbol{R}^{r}$, where $r$
is the rank of $V_{n}$. Define, $\mathcal{D}$ to be the set of $y$'s such
that (\ref{eq:1}) holds for some $x\in \boldsymbol{R}^{p-r}$. Also, given $%
y\in \mathcal{D}$, let $\Omega _{y}$ be the set of $x$'s such that (\ref%
{eq:1}) holds. Those are,%
\begin{eqnarray}
\mathcal{D} &:&\mathcal{=}\left\{ y\in \boldsymbol{R}^{r}|\exists x\in 
\boldsymbol{R}^{p-r}:N\left( V_{n}\right) x+N^{\bot }\left( V_{n}\right)
y\in \mathbb{P}\right\} ,  \label{eq:8} \\
\Omega _{y} &:&=\left\{ x\in \boldsymbol{R}^{p-r}|N\left( V_{n}\right)
x+N^{\bot }\left( V_{n}\right) y\in \mathbb{P}\right\} .  \label{eq:9}
\end{eqnarray}%
We solve (\ref{eq:05}) in the next theorem.

\begin{theorem}
\label{thm:02}The optimal value in (\ref{eq:05}) is given by 
\begin{equation}
\rho _{NL}^{o}=\frac{1}{2}\max_{i}\max_{y\in \mathcal{D}}\left[ \max_{x\in
\Omega _{y}}\mathcal{R}\left[ \bar{H}\right] _{i}x-\min_{x\in \Omega _{y}}%
\mathcal{R}\left[ \bar{H}\right] _{i}x\right] ,  \label{optimal_NL}
\end{equation}%
where $\bar{H}=H_{n}N\left( V_{n}\right) $.
\end{theorem}

The above theorem characterizes the optimal error when the MCF is not
restricted to any particular class. Furthermore, one can cast (\ref%
{optimal_NL}) as a linear program and hence compute it in a tractable way.
In fact, the optimal cost in (\ref{optimal_NL}) can be rewritten as%
\begin{equation*}
\rho _{NL}^{o}=\frac{1}{2}\min \bar{\eta},
\end{equation*}%
subject to%
\begin{eqnarray*}
-\eta _{i} &\leq &\bar{\eta}, \\
-\mathcal{R}\left[ \bar{H}\right] _{i}x_{1}+\mathcal{R}\left[ \bar{H}\right]
_{i}x_{2} &\leq &\eta _{i}, \\
N\left( V_{n}\right) x_{j}+N^{\bot }\left( V_{n}\right) y &\geq &0, \\
\mathbf{1}^{T}N\left( V_{n}\right) x_{j}+\mathbf{1}^{T}N^{\bot }\left(
V_{n}\right) y &=&1,
\end{eqnarray*}%
for all $i=1,2,...,m$ and $j=1,2$, where $m$ is the number of rows of $\bar{H%
}$. However, computing the MCF itself is a harder problem. In fact, the
optimal moment closure function is parametrized by $y\in \mathcal{D}$ and is
constructed in the proof of the above theorem. It is given by (\ref{eq:opt1}%
) and (\ref{eq:opt2}) , and can be computed via LP for a given value of $%
y\in \mathcal{D}$. However, as the LPs do not have a closed form, $\phi
^{optimal}\left( .\right) $ does not have a closed form either and this
makes the use of this MCF challenging from the computational point of view.
Therefore, we focus on the affine moment closure functions next and show
that designing the optimal affine MCF is in fact a LP and hence tractable.
Moreover, we compare the performance of the affine MCF (defined in (\ref%
{eq:20})) with (\ref{optimal_NL}) and show that nonlinear MCF cannot
outperform affine ones.

\subsection{Affine versus Nonlinear Moment Closure Functions}

In this section, we consider affine moment closure functions of the form%
\begin{equation}
\phi _{Affine}\left( V_{n}P\right) =KV_{n}P+K_{0},  \label{eq:20}
\end{equation}%
where $K$ and $K_{0}$ are matrices with appropriate dimensions. For
compactness, we adopt the following notation:%
\begin{eqnarray*}
\rho _{affine}\left( K,K_{0}\right) &=&\sup_{P\in \mathbb{P}}\left\Vert
H_{n}P-\left( KV_{n}P+K_{0}\right) \right\Vert , \\
\rho _{affine}^{o} &=&\inf_{K,K_{0}}\sup_{P\in \mathbb{P}}\left\Vert
H_{n}P-\left( KV_{n}P+K_{0}\right) \right\Vert .
\end{eqnarray*}

We note that given $K$ and $K_{0}$, one can use Lemma \ref{lemma:01} to
compute $\rho _{affine}\left( K,K_{0}\right) $ as 
\begin{equation*}
\rho _{affine}\left( K,K_{0}\right) =\left\Vert H_{n}-\left( KV_{n}+K_{0}%
\mathbf{1}^{T}\right) \right\Vert _{l_{1}-l_{\infty }},
\end{equation*}%
where we used $K_{0}=K_{0}\mathbf{1}^{T}P$ for $P\in \mathbb{P}$.

The next theorem provides a LP for computing the optimal affine moment
closure function.

\bigskip

\begin{theorem}
\label{theorem:AffineMCF}The optimal affine moment closure function, in the
form (\ref{eq:20}), can be found from the following LP:%
\begin{equation*}
\rho _{affine}^{o}=\inf_{\phi \text{ affine}}\sup_{P\in \mathbb{P}}\rho
\left( P,\phi \right) =\min_{K,K_{0}}\gamma 
\end{equation*}%
subject to%
\begin{equation*}
-\gamma \mathbf{1}^{T}\leq \mathcal{R}\left[ H_{n}-\left( KV_{n}+K_{0}%
\mathbf{1}^{T}\right) \right] _{i}\leq \gamma \mathbf{1}^{T},
\end{equation*}%
for $i=1,2,...,m$, where $m$ is the number of rows in $H_{n}$. Furthermore,
the optimal cost is given by 
\begin{equation}
\rho _{affine}^{o}=\inf_{\phi \text{ affine}}\sup_{P\in \mathbb{P}}\rho
\left( P,\phi \right) =\max_{i\in \left\{ 1,2,...,m\right\} }\max_{f}\left[ 
\mathcal{R}\left[ H_{n}\right] _{i}f\right] ,  \label{eq:40}
\end{equation}%
subject to%
\begin{eqnarray}
\left[ 
\begin{array}{c}
V_{n} \\ 
\mathbf{1}^{T}%
\end{array}%
\right] f &=&0,  \label{eq:35} \\
\left\Vert f\right\Vert _{l_{1}} &\leq &1.  \label{eq:36}
\end{eqnarray}
\end{theorem}

The above theorem provides the machinery to find the optimal affine moment
closure function. In general, one expects that $\rho _{affine}^{o}\geq \rho
_{NL}^{o}$, as affine functions form a proper subset of all functions.
However, in what follows, we will show that no moment closure function can
outperform affine ones.

\begin{theorem}
\label{theorem:NL vs Affine}The following equality holds%
\begin{equation}
\rho _{NL}^{o}=\rho _{affine}^{o}.  \label{eq:06}
\end{equation}
\end{theorem}

In light of this theorem, we use the affine MCF in our RMC scheme. Next, we
quantify the error between the true and the approximate system of moments.

\subsection{Error Quantification}

We derive the error bounds between the closed system of moments (\ref{eq:11}%
) and the true system (\ref{eq:03}) by studying the error dynamics. Let the
error be given by $e$, where $e=E_{n}-\nu _{n}$. Then, the error dynamics is
given by%
\begin{eqnarray*}
\dot{e} &=&AE_{n}+b\mu _{n+1}-A\nu _{n}-bK\nu _{n}-bK_{0} \\
&=&\left( A+bK\right) e+b\left[ \mu _{n+1}-\left( KE_{n}+K_{0}\right) \right]
;e\left( 0\right) =0.
\end{eqnarray*}%
Furthermore, for $i=1,2,...,n$, $e_{i}$, which is the error in the $i^{th}$
moment between the true and the closed system, can be written as%
\begin{equation}
e_{i}\left( t\right) =\int_{0}^{t}c_{i}\Phi \left( t,\tau \right) \left[ \mu
_{n+1}\left( \tau \right) -\left( KE_{n}\left( \tau \right) +K_{0}\right) %
\right] d\tau ,  \label{eq:09}
\end{equation}%
where $c_{i}$ is defined in (\ref{eq:ci}); $\Phi \left( t,\tau \right) $ is
the state transition matrix associated with the pair $\left( A+bK,b\right) $
and%
\begin{equation*}
\frac{d}{dt}\Phi \left( t,t_{0}\right) =\left( A\left( t\right) +b\left(
t\right) K\right) \Phi \left( t,t_{0}\right) ;\text{ with }\Phi \left(
t_{0},t_{0}\right) =I.
\end{equation*}%
This error is quantified in the next theorem.

\begin{theorem}
\label{thm:01}Given $K$ and $K_{0}$, the error in the $i^{th}$ moment due to
the RMC is given by%
\begin{equation}
ess\sup_{t\in \left[ 0,T\right] }\left\Vert e_{i}\left( t\right) \right\Vert
\leq \left[ \int_{0}^{T}\left\Vert c_{i}\Phi \left( t,\tau \right)
b\right\Vert dt\right] \times \rho _{affine}\left( K,K_{0}\right) .
\label{eq:10}
\end{equation}
\end{theorem}

\section{Conclusion}

In this paper, we studied the moment closure problem for the CME. We
developed the Robust Moment Closure technique in which we used the affine
moment closure functions to approximate the higher order moments in terms of
the lower order ones. We showed that, in the absence of a priori information
on the probability distribution, the affine MCFs are optimal and,
furthermore, they can be found via LP. Consequently, utilizing the affine
moment closure functions, we derived a system of finite dimension that
approximates the low-order moments. Moreover, we quantified the error in
this approximation in terms of the $l_{\infty }$ induced norm of some linear
system. Our results allow for the explicit simulation and analytical
computation of approximate moments, which can be effectively used to
characterize the noise properties of the biochemical network under study.

\section*{APPENDIX}

\subsection*{Proof on Theorem \protect\ref{thm:02}}

First, notice that, the definition of $l_{\infty }$norm, (\ref{eq:05}) can
be rewritten as%
\begin{equation}
\rho _{NL}^{o}=\inf_{\phi _{1},\phi _{2},...,\phi _{m}}\max_{i\in \left\{
1,2,...,m\right\} }\sup_{P\in \mathbb{P}^{p}}\left\vert \mathcal{R}\left[
H_{n}\right] _{i}P-\phi _{i}\left( V_{n}P\right) \right\vert ,
\end{equation}%
where $\phi _{i}\left( .\right) $ is the $i^{th}$ entry of vector $\phi
\left( .\right) $ and $\left\vert .\right\vert $ is the absolute value
function. Hence, $\rho _{NL}^{o}=\max \left\{ \bar{\eta}_{1},\bar{\eta}%
_{2},...,\bar{\eta}_{m}\right\} $, where%
\begin{equation}
\bar{\eta}_{i}=\inf_{\phi _{i}}\sup_{P\in \mathbb{P}^{p}}\left\vert \mathcal{%
R}\left[ H_{n}\right] _{i}P-\phi _{i}\left( V_{n}P\right) \right\vert ,\text{
for }i=1,2,...,m.  \label{eq:6}
\end{equation}%
Given $i\in \left\{ 1,2,...,m\right\} $, (\ref{eq:06}) is equivalent to%
\begin{eqnarray}
&&\bar{\eta}_{i}=\inf_{\phi _{i}}\max_{y\in \mathcal{D}}\max_{{\normalsize %
x\in \Omega }_{y}}  \label{eq:7} \\
&&\left\vert \mathcal{R}\left[ H_{n}\right] _{i}\left( N\left( V_{n}\right)
x+N^{\bot }\left( V_{n}\right) y\right) -\phi _{i}\left( V_{n}N^{\bot
}\left( V_{n}\right) y\right) \right\vert ,
\end{eqnarray}%
where we used (\ref{eq:1}), and $\mathcal{D}$ and $\Omega _{y}$ are defined
in (\ref{eq:8})-(\ref{eq:9}). Since, $\phi _{i}\left( .\right) $ is not
restricted to any class of functions, one can define 
\begin{equation}
f_{i}\left( y\right) =\phi _{i}\left( V_{n}N^{\bot }\left( V_{n}\right)
y\right) -\mathcal{R}\left[ H_{n}\right] _{i}N^{\bot }\left( V_{n}\right) y,
\label{eq:72}
\end{equation}%
and perform the optimization over $f_{i}\left( y\right) $. Therefore, (\ref%
{eq:7}) reduces to%
\begin{equation}
\bar{\eta}_{i}=\inf_{f_{i}}\max_{y\in \mathcal{D}}\max_{{\normalsize x\in
\Omega }_{y}}\left\vert \mathcal{R}\left[ \bar{H}\right] _{i}x-f_{i}\left(
y\right) \right\vert ,  \label{eq:70}
\end{equation}%
where $\bar{H}=H_{n}N\left( V_{n}\right) $. From (\ref{eq:70}), we have 
\begin{equation}
\bar{\eta}_{i}\geq \max_{y\in \mathcal{D}}\min_{f_{i}\left( y\right) }\max_{%
{\normalsize x\in \Omega }_{y}}\left\vert \mathcal{R}\left[ \bar{H}\right]
_{i}x-f_{i}\left( y\right) \right\vert .  \label{eq:71}
\end{equation}%
Now, given $y\in \mathcal{D}$, the optimal value of $f_{i}\left( y\right) $
to minimize $\max_{{\normalsize x\in \Omega }_{y}}\left\vert \mathcal{R}%
\left[ \bar{H}\right] _{i}x-f_{i}\left( y\right) \right\vert $ is the
algebraic average between the largest and smallest values of $\mathcal{R}%
\left[ \bar{H}\right] _{i}x$ where $x\in \Omega _{y}$. That is,%
\begin{equation}
f_{i}^{optimal}\left( y\right) =\frac{1}{2}\left[ \max_{x_{1}{\normalsize %
\in \Omega }_{y}}\mathcal{R}\left[ \bar{H}\right] _{i}x_{1}+\min_{x_{2}%
{\normalsize \in \Omega }_{y}}\mathcal{R}\left[ \bar{H}\right] _{i}x_{2}%
\right] .  \label{eq:opt1}
\end{equation}%
In this case,%
\begin{eqnarray*}
&&\min_{f_{i}\left( y\right) }\max_{x{\normalsize \in \Omega }%
_{y}}\left\vert \mathcal{R}\left[ \bar{H}\right] _{i}x-f_{i}\left( y\right)
\right\vert \\
&=&\frac{1}{2}\left[ \max_{x_{1}{\normalsize \in \Omega }_{y}}\mathcal{R}%
\left[ \bar{H}\right] _{i}x_{1}-\min_{x_{2}{\normalsize \in \Omega }_{y}}%
\mathcal{R}\left[ \bar{H}\right] _{i}x_{2}\right] \\
&=&-\frac{1}{2}\left[ \min_{x_{1}{\normalsize \in \Omega }_{y}}-\mathcal{R}%
\left[ \bar{H}\right] _{i}x_{1}+\min_{x_{2}{\normalsize \in \Omega }_{y}}%
\mathcal{R}\left[ \bar{H}\right] _{i}x_{2}\right]
\end{eqnarray*}%
Therefore, the lower bound in (\ref{eq:71}) is given by%
\begin{equation}
\bar{\eta}_{i}\geq \frac{1}{2}\max_{y\in \mathcal{D}}\left[ \max_{x%
{\normalsize \in \Omega }_{y}}\mathcal{R}\left[ \bar{H}\right] _{i}x-\min_{x%
{\normalsize \in \Omega }_{y}}\mathcal{R}\left[ \bar{H}\right] _{i}x\right] .
\label{eq:3}
\end{equation}%
Furthermore, the lower bound, in (\ref{eq:71}), is achievable. In fact, the
lower bound is attainable for any $f_{i}\left( .\right) $, in (\ref{eq:70}),
with the property that it coincides with $f_{i}^{optimal}\left( .\right) $
on the set $\mathcal{D}$. From (\ref{eq:72}), an optimal $\phi
_{i}^{optimal}\left( .\right) $ is the one whose value at $V_{n}N^{\bot
}\left( V_{n}\right) y$, for $y\in \mathcal{D}$, is given by 
\begin{equation}
\phi _{i}^{optimal}\left( V_{n}N^{\bot }\left( V_{n}\right) y\right)
=f_{i}^{optimal}\left( y\right) +\mathcal{R}\left[ H_{n}\right] _{i}N^{\bot
}\left( V_{n}\right) y,  \label{eq:opt2}
\end{equation}%
and arbitrary otherwise. That is (\ref{eq:3}) is a tight inequality and
hence taking the $\max $ over $i$ completes the proof.

\subsection*{Proof of Theorem \protect\ref{theorem:AffineMCF}}

Notice that for affine moment closure functions we have%
\begin{eqnarray}
&&\rho _{affine}^{o}=\min_{K,K_{0}}\max_{P\in \mathbb{P}}\left\Vert
H_{n}P-\left( KV_{n}P+K_{0}\right) \right\Vert  \notag \\
&=&\min_{K,K_{0}}\max_{P\in \mathbb{P}}\left\Vert \left[ H_{n}-\left(
KV_{n}+K_{0}\mathbf{1}^{T}\right) \right] P\right\Vert .  \label{eq:21}
\end{eqnarray}%
By Lemma \ref{lemma:01}, one obtains%
\begin{equation}
\rho _{affine}^{o}=\min_{K,K_{0}}\max_{i,j}\left\vert \left[ H_{n}-\left(
KV_{n}+K_{0}\mathbf{1}^{T}\right) \right] _{ij}\right\vert ,  \label{eq:22}
\end{equation}%
where $\left[ H_{n}-\left( KV_{n}+K_{0}\mathbf{1}^{T}\right) \right] _{ij}$
denotes the entry on row $i$ and column $j$ of the matrix $H_{n}-\left(
KV_{n}+K_{0}\mathbf{1}^{T}\right) \in \mathbb{R}^{m\times p}$. Then,%
\begin{equation*}
\max_{i,j}\left\Vert \left[ H_{n}-\left( KV_{n}+K_{0}\mathbf{1}^{T}\right) %
\right] _{ij}\right\Vert =\min \gamma ,
\end{equation*}%
subject to 
\begin{equation*}
-\gamma \mathbf{1}^{T}\leq \mathcal{R}\left[ H_{n}-\left( KV_{n}+K_{0}%
\mathbf{1}^{T}\right) \right] _{i}\leq \gamma \mathbf{1}^{T},
\end{equation*}
for $i=1,2,...,m$. This completes the proof of the first part. For the
second part, note that (\ref{eq:22}) can be written as%
\begin{eqnarray*}
&&\rho _{affine}^{o}=\min_{\substack{ \gamma \geq 0  \\ K,K_{0}}}%
\max_{i}\max_{\zeta _{i}\in \mathbb{R}_{\geq 0}^{p},\xi _{i}\in \mathbb{R}%
_{\geq 0}^{p}} \\
&&\gamma +\left( \mathcal{R}\left[ H_{n}-\left( KV_{n}+K_{0}\mathbf{1}%
^{T}\right) \right] _{i}-\gamma \mathbf{1}^{T}\right) \zeta _{i} \\
&&-\left( \mathcal{R}\left[ H_{n}-\left( KV_{n}+K_{0}\mathbf{1}^{T}\right) %
\right] _{i}+\gamma \mathbf{1}^{T}\right) \xi _{i},
\end{eqnarray*}%
where $\zeta _{i}$'s and $\xi _{i}$'s are the Lagrange multipliers. Due to
the convexity of the objective function and constraints (zero duality gap),
one can change the order of the $\min $ and $\max $. Hence,%
\begin{equation}
\rho _{affine}^{o}=\max_{i}\max_{\zeta _{i}\in \mathbb{R}_{\geq 0}^{p},\xi
_{i}\in \mathbb{R}_{\geq 0}^{p}}\mathcal{G}\left( \zeta ,\xi \right) ,
\label{eq:33}
\end{equation}%
where $\mathcal{G}\left( \zeta ,\xi \right) $ is the so-called Lagrangian
and is given by%
\begin{align*}
& \mathcal{G}_{i}\left( \zeta ,\xi \right) =\min_{\substack{ \gamma \geq 0 
\\ K,K_{0}}}\left[ 1-\mathbf{1}^{T}\sum_{i=1}^{m}\left( \zeta _{i}+\xi
_{i}\right) \right] \gamma \\
& +\left[ \mathcal{R}\left[ H_{n}\right] _{i}-\mathcal{R}\left[ K\right]
_{i}V_{n}-\mathcal{R}\left[ K_{0}\right] _{i}\mathbf{1}^{T}\right] \left(
\zeta _{i}-\xi _{i}\right) .
\end{align*}%
One can easily verify that 
\begin{equation}
\mathcal{G}_{i}\left( \zeta ,\xi \right) =\mathcal{R}\left[ H_{n}\right]
_{i}\left( \zeta _{i}-\xi _{i}\right) ,  \label{eq:34}
\end{equation}%
if 
\begin{eqnarray}
V_{n}\left( \zeta _{i}-\xi _{i}\right) &=&0,\text{ for }i\in \left\{
1,2,...,m\right\} ,  \label{eq:30} \\
\mathbf{1}^{T}\left( \zeta _{i}-\xi _{i}\right) &=&0,\text{ for }i\in
\left\{ 1,2,...,m\right\} ,  \label{eq:31} \\
\mathbf{1}^{T}\left( \zeta _{i}+\xi _{i}\right) &\leq &1,  \label{eq:32}
\end{eqnarray}%
and otherwise $\mathcal{G}_{i}\left( \zeta ,\xi \right) =-\infty $. The
proof is complete by defining the new variables $f_{i}=\zeta _{i}-\xi _{i}$.
More precisely, for any set of $\zeta _{i}$'s and $\xi _{i}$'s that
satisfies (\ref{eq:30})-(\ref{eq:32}), one can define $f_{i}=\zeta _{i}-\xi
_{i}$ satisfying (\ref{eq:35})-(\ref{eq:36}). And, conversely, for any set
of $f_{i}$'s that satisfies (\ref{eq:35}) and (\ref{eq:36}), define $\zeta
_{i}=f_{i}^{+}$ and $\xi _{i}=f_{i}^{-}$ where $f_{i}=f_{i}^{+}-f_{i}^{-}$
is the positive decomposition of $f_{i}$.

\subsection*{Proof of Theorem \protect\ref{theorem:NL vs Affine}}

We will show that $\rho _{NL}^{o}\geq \rho _{affine}^{o}$, where $\rho
_{NL}^{o}$ and $\rho _{affine}^{o}$ are given in (\ref{optimal_NL}) and (\ref%
{eq:40}), respectively. To this end, let $i^{\ast }$ and $f^{\ast }$ be the
maximizers of (\ref{eq:40}). That is, 
\begin{equation}
V_{n}f^{\ast }=0,\mathbf{1}^{T}f^{\ast }=0,\left\Vert f^{\ast }\right\Vert
=1,  \label{eq:102}
\end{equation}%
and $\rho _{affine}^{o}=\mathcal{R}\left[ H\right] _{i^{\ast }}f^{\ast }$.
Now, let $f^{\ast }=\left( f^{\ast }\right) ^{+}-\left( f^{\ast }\right)
^{-} $ be the positive decomposition of $f^{\ast }$. The nonnegative vectors 
$\left( f^{\ast }\right) ^{+}$ and $\left( f^{\ast }\right) ^{-}$ can be
written as unique summations of elements from $\mathcal{N}\left(
V_{n}\right) $ and $\mathcal{N}^{\bot }\left( V_{n}\right) $. More
precisely, there exist $\alpha _{1},\alpha _{2}\in \mathbb{R}^{q-r}$ and $%
\beta _{1},\beta _{2}\in \mathbb{R}^{r}$ such that%
\begin{eqnarray}
\left( f^{\ast }\right) ^{+} &=&N\left( V_{n}\right) \alpha _{1}+N^{\bot
}\left( V_{n}\right) \beta _{1}, \\
\left( f^{\ast }\right) ^{-} &=&N\left( V_{n}\right) \alpha _{2}+N^{\bot
}\left( V_{n}\right) \beta _{2}.
\end{eqnarray}%
Furthermore, from (\ref{eq:102}) we have%
\begin{eqnarray*}
&&V_{n}\left( f^{\ast }\right) ^{+}-V_{n}\left( f^{\ast }\right)
^{-}=V_{n}N^{\bot }\left( V_{n}\right) \left( \beta _{1}-\beta _{2}\right)
=0, \\
&&\mathbf{1}^{T}N\left( V_{n}\right) \left( \alpha _{1}-\alpha _{2}\right) +%
\mathbf{1}^{T}N^{\bot }\left( V_{n}\right) \left( \beta _{1}-\beta
_{2}\right) =0, \\
&&\mathbf{1}^{T}N\left( V_{n}\right) \left( \alpha _{1}+\alpha _{2}\right) +%
\mathbf{1}^{T}N^{\bot }\left( V_{n}\right) \left( \beta _{1}+\beta
_{2}\right) =1.
\end{eqnarray*}%
From above expressions, since $V_{n}N^{\bot }\left( V_{n}\right) $ is full
column rank, $\beta _{1}=\beta _{2}=\beta $, $\mathbf{1}^{T}N\left(
V_{n}\right) \alpha _{1}=\mathbf{1}^{T}N\left( V_{n}\right) \alpha _{2}$,
and 
\begin{eqnarray*}
2\mathbf{1}^{T}N\left( V_{n}\right) \alpha _{1}+2\mathbf{1}^{T}N^{\bot
}\left( V_{n}\right) \left( \beta \right) &=&1, \\
2\mathbf{1}^{T}N\left( V_{n}\right) \alpha _{2}+2\mathbf{1}^{T}N^{\bot
}\left( V_{n}\right) \left( \beta \right) &=&1.
\end{eqnarray*}%
Let $x_{1}=2\alpha _{1}$, $x_{2}=2\alpha _{2}$, and $y=2\beta $. Then, it is
easy to verify that $y\in \mathcal{D}$ and $x_{1},x_{2}\in \Omega _{y}$,
where $\mathcal{D}$ and $\Omega _{y}$ are defined in (\ref{eq:8})-(\ref{eq:9}%
). Also, from (\ref{optimal_NL}), we have 
\begin{eqnarray*}
\rho _{NL}^{o} &=&\frac{1}{2}\max_{i}\max_{y\in \mathcal{D}}\left[
\max_{x\in \Omega _{y}}\mathcal{R}\left[ \bar{H}\right] _{i}x+\max_{x\in
\Omega _{y}}-\mathcal{R}\left[ \bar{H}\right] _{i}x\right] \\
&\geq &\frac{1}{2}\left[ \mathcal{R}\left[ \bar{H}\right] _{i^{\ast }}x_{1}-%
\mathcal{R}\left[ \bar{H}\right] _{i^{\ast }}x_{2}\right] \\
&=&\left[ \mathcal{R}\left[ H\right] _{i^{\ast }}N\left( V_{n}\right) \alpha
_{1}-\mathcal{R}\left[ H\right] _{i^{\ast }}N\left( V_{n}\right) \alpha _{2}%
\right] \\
&=&\mathcal{R}\left[ H\right] _{i^{\ast }}\left[ \left( f^{\ast }\right)
^{+}-\left( f^{\ast }\right) ^{-}\right] =\mathcal{R}\left[ H\right]
_{i^{\ast }}f^{\ast }=\rho _{affine}^{o}.
\end{eqnarray*}%
The proof is complete.

\subsection*{Proof of Theorem \protect\ref{thm:01}}

The proof is based on the direct calculation; from (\ref{eq:09}), one obtains%
\begin{eqnarray*}
&&ess\sup_{t\in \left[ 0,T\right] }\left\Vert e_{i}\left( t\right)
\right\Vert =\sup_{t\in \left[ 0,T\right] } \\
&&\left\Vert \int_{0}^{t}c_{i}\Phi \left( t,\tau \right) b\left[ \mu
_{n+1}\left( \tau \right) -\left( KE_{n}\left( \tau \right) +K_{0}\right) %
\right] d\tau \right\Vert \\
&\leq &\sup_{t\in \left[ 0,T\right] }\int_{0}^{t}\left\Vert c_{i}\Phi \left(
t,\tau \right) b\right\Vert \bar{\rho}_{L}\left( K,K_{0}\right) d\tau ,
\end{eqnarray*}%
which is the same is (\ref{eq:10}).

\bibliographystyle{IEEEtran}
\bibliography{BIBMC}

\end{document}